\documentclass[letterpaper, 10pt, conference]{IEEEtran}

\pdfoutput=1

\usepackage[
  backend=biber,
  bibstyle=ieee,
  citestyle=numeric-comp,
  sorting=nyt,
  mincitenames=1,
  maxcitenames=1,
]{biblatex} 
\DeclareCiteCommand{\citeauthor}
  {\boolfalse{citetracker}%
   \boolfalse{pagetracker}%
   \usebibmacro{prenote}}
  {\ifciteindex
     {\indexnames{labelname}}
     {}%
   \printtext[bibhyperref]{\printnames{labelname}}}
  {\multicitedelim}
  {\usebibmacro{postnote}}

\usepackage{flushend}
\usepackage{hyperref}
\usepackage{inconsolata}
\usepackage[
  activate={true, nocompatibility},
  final,
  kerning=true,
  spacing=true,
  factor=1100,
]{microtype}

\IEEEoverridecommandlockouts
\IEEEpubid{%
  \makebox[\columnwidth]{%
    978-1-6654-4905-2/21/\$31.00 \copyright 2021 IEEE%
    \hfill%
  }%
  \hspace{\columnsep}%
  \makebox[\columnwidth]{ }%
}

\begin{document}

\title{Do Abstractions Have Politics?\\Toward a More Critical Algorithm Analysis}

\author{%
  \IEEEauthorblockN{Kevin Lin}
  \IEEEauthorblockA{%
    \textit{Paul G. Allen School of Computer Science \& Engineering}\\%
    \textit{University of Washington}\\%
    Seattle, WA, USA\\%
    \href{mailto:kevinl@cs.uw.edu}{kevinl@cs.uw.edu}%
  }%
}

\maketitle

\begin{abstract}
The expansion of computer science (CS) education in K--12 and higher-education in the United States has prompted deeper engagement with equity that moves beyond inclusion toward a more critical CS education. Rather than frame computing as a value-neutral tool, a justice-centered approach to equitable CS education draws on critical pedagogy to ensure the rightful presence of political struggles by emphasizing the development of not only knowledge and skills but also CS disciplinary identities. While recent efforts have integrated ethics into several areas of the undergraduate CS curriculum, critical approaches for teaching data structures and algorithms in particular are undertheorized. Basic Data Structures remains focused on runtime-centered algorithm analysis.

We argue for affordance analysis, a more critical algorithm analysis based on an affordance account of value embedding. Drawing on critical methods from science and technology studies, philosophy of technology, and human-computer interaction, affordance analysis examines how the design of computational abstractions such as data structures and algorithms embody affordances, which in turn embody values with political consequences. We illustrate 5 case studies of how affordance analysis refutes social determination of technology, foregrounds the limitations of data abstractions, and implicates the design of algorithms in disproportionately distributing benefits and harms to particular social identities within the matrix of domination.
\end{abstract}


Current discourse around CSforAll and broadening participation in computing frame ``equity as inclusion'' \cite{calabrese2020beyond, vakil2018ethics}, calling for an extension of access to a high-quality computing education as a fundamental right for all students. ``However, are we satisfied with everyone learning to code, if the end game is to produce (admittedly more `diverse') coders who will primarily work to ensure the continued profitability of capitalist start-ups and technology giants?'' \cite{costanza2020design}. Computing education has traditionally positioned computing as value-neutral and more interested in efficiency and business profit than ``do[ing] something good'' \cite{vakil2018ethics, vakil2020ve, ko2020time, malazita2019infrastructures}. This approach reinforces the dominant narratives about the apolitical disciplinary identity of computer science and reproduces systems of oppression that dehumanize and invalidate marginalized students' experiences and perspectives \cite{calabrese2020beyond, vakil2020ve, agarwal2019integrating, ryoo2020take, malazita2019infrastructures}.

Technologies embody social relations and political power \cite{costanza2020design, winner1980artifacts, klenk2020technological}. A more equitable computer science (CS) education thus requires a more critical CS education that ``recognizes computing is not an unequivocal social good'' \cite{kafai2019theory} and centers the ``rightful presence'' of political struggles: the ``fraught histories'' and ``concrete injustices'' experienced by students in the computing classroom \cite{calabrese2020beyond, ryoo2020take}. ``To move toward a justice-centered approach to equity, \citeauthor{vakil2018ethics} argues, we must simultaneously attend to at least three features of CS education: the content of curriculum, the design of learning environments, and the politics and purposes of CS education reform.'' Specifically, \citeauthor{ko2020time} define three ideas for a more critical CS education: that computing has limits, data has limits, and CS has responsibility.

Recent efforts to design a more critical CS education in higher education include standalone ethics courses \cite{reich2020teaching, ferreira2021deep}; ethics integrated across the undergraduate computing curriculum \cite{grosz2019embedded, cohen2021new}; and integrated ethics in specific courses such as machine learning \cite{saltz2019integrating}, human-centered computing \cite{skirpan2018ethics}, and introductory CS \cite{fiesler2021intro, doore2020assignments}. In emphasizing the unjust distribution of benefits and harms caused by algorithmic decision-making systems, these efforts reflect a recent turn toward a structural and systemic analysis of computing injustices that ``explore ethics in relation to institutions, societies, ideology, or epistemological perspectives in CS rather than a focus on the `good' and `bad' decisions individual actors make in their interactions with technology'' \cite{vakil2018ethics}.

``Both [the machine learning and mechanism design communities] have been heeding the call for attention to values, politics, and `social good' more generally, holding more than twenty technical workshops and conferences between them on some variation of fairness, bias, discrimination, accountability, and transparency in the last five years'' \cite{abebe2020roles}. Similarly, social scientists have begun studying algorithms as opaque ``black boxes'' that require ``a range of methodological strategies in order to bypass these layers of impenetrability and document the inner working of computational systems'' \cite{christin2020ethnographer}. However, it is less clear how these efforts might translate to data structures and algorithms: the study of computational abstractions underpinning such systems. Critically, methods that treat systems as black boxes will not implicate the design of data structures and algorithms toward the system's decision-making, values, and outcomes. Far from studying a black box system, the study of data structures and algorithms is uniquely positioned in the undergraduate CS curriculum to enable critical examination of the inner workings of computational systems in order to address the question, ``Do abstractions have politics?''

\section*{Asymptotic Analysis}

A survey and expert panel study conducted by \citeauthor{porter2018developing} identified the following learning goals for Basic Data Structures.
\begin{enumerate}
  \small
  \item Analyze runtime efficiency of algorithms related to data structure design.
  \item Select appropriate abstract data types for use in a given application.
  \item Compare data structure tradeoffs to select the appropriate implementation for an abstract data type.
  \item Design and modify data structures capable of insertion, deletion, search, and related operations.
  \item Trace through and predict the behavior of algorithms (including code) designed to implement data structure operations.
  \item Identify and remedy flaws in a data structure implementation that may cause its behavior to differ from the intended design.
\end{enumerate}
These learning goals present a unique challenge for designing a more critical data structures and algorithms course. ``Abstract data types were introduced as a way of freeing a programmer from concern about irrelevant details in his use of data abstractions'' \cite{liskov1974programming}, divorcing information---and the values embodied therein---from the data structures and algorithms that organize them \cite{malazita2019infrastructures}. Rather than examine the applications of impacts data structures and algorithms on society, Basic Data Structures focuses on examining their efficiency: algorithm analysis is defined as ``[r]untime analysis and/or space complexity'' \cite{porter2018developing} using asymptotic notation, such as Big O notation.

Algorithm design and implementation is thus a means of realizing a specification or abstract data type without critically questioning the design of the abstraction \cite{costanza2020design, malazita2019infrastructures}. Design issues and tradeoffs are narrowly framed in terms of code behavior and program efficiency rather than design justice---``design that is led by marginalized communities and that aims explicitly to challenge, rather than reproduce, structural inequalities'' \cite{costanza2020design}. The rightful presence of political struggles in the data structures and algorithms classroom is predicated on a more critical algorithm analysis that examines the values embodied in data structures and algorithms (and applications thereof).

\section*{Affordance Theory}

Affordance analysis is an alternative algorithm analysis that draws on science and technology studies, philosophy of technology, and human-computer interaction to examine how computational abstractions such as data structures and algorithms embody affordances. Affordances are relational properties of objects that make ``specific outcomes more likely given the circumstances provided that the subject aims to bring about these outcomes'' \cite{klenk2020technological}. For example, ``a chair \emph{affords} sitting, a doorknob \emph{affords} turning, a mouse \emph{affords} moving the cursor on the screen and clicking at a particular location, and a touchscreen \emph{affords} tapping and swiping'' \cite{costanza2020design}.

The \emph{affordance account of value embedding} is a theory for understanding how technological artifacts embody moral values \cite{klenk2020technological}. Affordances are not value-neutral: like the disposition to act in a certain way, affordances enable outcomes that may be valuable. An artifact such as an assault rifle has ``negative value because it enables killing in a broad range of circumstances'' \cite{klenk2020technological}. In general, the value of an artifact is determined by the ``actions or events it affords'' and their resulting values \cite{klenk2020technological}.

\section*{Affordance Analysis}

Affordance analysis applies an affordance account of value embedding toward computational abstractions such as data structures and algorithms. If these abstractions embody affordances that in turn embody values, abstractions can produce ``consequences logically and temporally \emph{prior} to any of its professed uses'' \cite{winner1980artifacts}. In other words, abstractions have politics through the values embodied by their affordances.
\begin{quote}
    \small
    For engineers aiming to design for value, the affordance account already indicates a simple recipe: measure which actions a given artefact makes likely given a context (for which we can build on social scientific tools) and then evaluate whether these affordances are legitimately desirable (for which we have normative ethics). \cite{klenk2020technological}
\end{quote}

To identify the affordances of a programming abstraction, consider its Application Programming Interface (API), which ``lists the affordances that a software entity makes available'' \cite{abbott2019bit}. Data structures often implement abstract data types that provide common programming interfaces \cite{liskov1974programming}. In Java, a \texttt{class} or \texttt{interface} defines public methods that afford certain actions.

To evaluate an affordance according to its effects on social systems and institutions \cite{costanza2020design, abebe2020roles, christin2020ethnographer, wong2020infrastructural}, consider \citeauthor{ferreira2021deep}.
\begin{LaTeXdescription}
  \small
  \item[History and Context] When examining a specific technology, what are the historical and cultural circumstances in which it emerged? When was it developed? For what purpose? How has its usage and function changed from then to today?
  \item[Power Dynamics and Hegemony] Who benefits from this technology? At the expense of whose labor? How is this technology sold and marketed? What are the economic and political interests for the proliferation of this technology?
  \item[Developing Effective Long-Term Solutions] What solutions are currently being implemented to address this labor/benefit asymmetry? In what ways do they reinforce or challenge the status quo? What are the long- and short-term implications of these solutions and who will benefit from them?
\end{LaTeXdescription}
\citeauthor{wong2020infrastructural} organizes these questions under ``\emph{infrastructural speculation}, a lens to center and unravel the lifeworlds of speculative designs,'' presenting 8 design tactics to focus on ``the `background' practices surrounding technologies beyond use, to think about the broad---yet differential---impacts of infrastructures and contend with questions of institutional power.'' By attending to ``the lifeworld of artifacts---the social, perceptual, and political environment in which they exist,'' \cite{wong2020infrastructural} affordance analysis implicates the affordances of computational abstractions such as data structures and algorithms to the systems they empower and the social futures they create.

To illustrate affordance analysis, we present 5 case studies for Basic Data Structures. More examples are available online.\footnote{\url{https://kevinl.info/do-abstractions-have-politics/}}

\subsection*{Priority Queues for Content Moderation}

A priority queue is an abstract data type where elements are retrieved according to their associated priority value. A max-oriented priority queue retrieves highest-priority elements first while a min-oriented priority queue retrieves lowest-priority elements first. Social media platforms rely on algorithms to draw attention to the most engaging user-generated content. In order to manage user-generated content, platforms design content moderation systems. A content moderation system might use a priority queue for human moderators to review flagged content by assigning the priority values according to the most toxic (severe, obscene, disrespectful, harmful, or otherwise disengaging) content.

Content moderation plays an understated but integral role in determining the content shown to users. Affordance analysis reveals that content moderation priority ordering embodies value. For users of the platform---particularly the most marginalized users---prioritizing moderation for the most toxic content may not necessarily reduce the most harmful content. Some users might consider personalized harassment or identity attacks as more harmful than toxic content identified by a general-purpose algorithm \cite{mahar2018squadbox}. For social media hackers who spread misinformation, the particular way in which content is prioritized can introduce loopholes with broad political impacts. Even if our priority values considered misinformation, we might further question whether misinformation should even be in the same priority queue as toxic content. Human moderators review hundreds of submissions everyday, leading to fatigue, mental health issues, and PTSD-like symptoms. Human moderators desensitized by repeated exposure to toxic content might find it harder to flag and remove misinformation.

A priority queue affords access to the highest/lowest priority-valued elements. Applied toward content moderation, priority queues optimize for review of certain content, distributing social power, benefits, and harms according to their priority values. By attending to the design of the priority queue abstraction, affordance analysis refutes \emph{social determination of technology}: the view that, ``What matters is not technology itself, but the social or economic system in which it is embedded'' \cite{winner1980artifacts}.

\subsection*{Binary Trees for Hiring Decisions}

Binary trees are the foundation for data structures such as binary search trees that implement associative sets and maps as well as binary heaps that implement priority queues. But beyond implementing abstract data types, binary trees can also directly represent hierarchical relationships between elements in the recursive tree structure such as in a decision tree. A hiring algorithm could represent its decision-making process as a binary tree. In this example, each internal node in the binary tree could represent a hiring question with a yes/no answer that corresponds to the left/right children, and each leaf node could represent a final yes/no hiring decision.

A binary tree affords questions that encode hard requirements for the hiring position because those questions can be answered with a yes/no answer, but other characteristics might be harder to represent. It might be hard to say exactly how much prior experience (or what kind of prior experience) is needed for the job beyond the core requirements, especially when there are many candidates with diverse backgrounds. Creating a more nuanced binary tree hiring decision-making system requires acknowledging this design constraint throughout the requirements planning and question design process. A design that affords yes/no question answers can preclude more open-ended questions that allow a broader diversity of ways for a candidate to demonstrate suitability, rather than only the most prevalent or dominant candidate experiences. By narrowly prioritizing design for the dominant candidate experiences while marginalizing others as edge cases, sociotechnical systems risk exacerbating social injustice.

It's possible to design a binary tree that is not limited to yes/no question answers. Just as we can rewrite multiway if/else-if/else conditionals into nested binary if/else conditionals, we can represent any multiway tree as a binary tree. But because binary trees afford binary questions, algorithms that rely on binary trees may tend toward solutions modeled with purely binary questions. Affordance analysis suggests that the decision to represent a hiring algorithm as a binary tree \emph{dis-affords} questions incompatible with binary answers.

\subsection*{Autocomplete for Search Engines}

Autocomplete is an application feature that helps a user select valid search results by showing possible inputs as they type. Wayne (in \citeauthor{parlante2016nifty}) describe \emph{Autocomplete-Me}, a simple autocompletion API designed as an assignment for Basic Data Structures. The Java API provides two key operations: a constructor that stores the corpus of all possible autocompletion terms and an \texttt{allMatches} method that returns all terms that start with the given prefix ordered by descending weight so that the most important terms appear first.

Our evaluation of these affordances might begin along the same lines as in \emph{Priority Queues for Content Moderation} by critiquing the weight (or importance) ranking. But the use of autocomplete in the specific context of search engines also draws attention to more structural issues. In \emph{Algorithms of Oppression}, \citeauthor{noble2018algorithms} implicates search engines in reproducing sexist, racist, or misogynistic ideas through their search suggestions and results. In 2013, ``[t]he Google Search autosuggestions featured a range of sexist ideas such as the following:
\begin{itemize}
    \small
    \item Women cannot: drive, be bishops, be trusted, speak in church
    \item Women should not: have rights, vote, work, box
    \item Women should: stay at home, be slaves, be in the kitchen, not speak in church
    \item Women need to: be put in their places, know their place, be controlled, be disciplined
\end{itemize}
These associations are exacerbated at the intersection of social identities. \citeauthor{noble2018algorithms} examined the racist and sexist ideas in the top autocompletion and search results for queries including ``Black girls'', ``Latinas'', and ``Asian girls''. With the ubiquity of search engines, such results embody cultural power: its impacts also extend to people who don't directly use Google Search. The value of a computational abstraction is not only in how it directly affects end users, but also how it affects \emph{societal systems and structures} by eroding civil and human rights.

\subsection*{Shortest Paths for Navigation Directions}

The single-pair shortest paths problem focuses on finding a shortest path between two nodes in a graph that minimizes the sum of the edge weights. Several well-known algorithms have been invented to solve variants of the shortest paths problem, including Dijkstra's algorithm and the A* search algorithm.

Shortest paths algorithms can be used to compute navigation directions for a mapping application. Consider a road network represented as an edge-weighted graph where junctions are nodes and edges are distances (or travel time) along the road segment between junctions. However, a shortest path is not necessarily the ``best'' path. It might not take into account mode of travel, road grade, or ability of the user. It might route through local roads not designed to sustain large amounts of traffic safely, increasing risks to the local neighborhood and other drivers. For pedestrian walking directions, especially at night, it might not offer the most well-lit or populated route. To address these problems, we could modify the edge weights to better match desired outcomes. But even so, the use of shortest paths algorithms for navigation directions might mask more structural issues such as public disinvestment municipal infrastructure and public transit.

Affordance analysis surfaces how algorithms can implicitly encode for an imagined default user situated atop the \emph{matrix of domination}: ``white, male, abled, English-speaking, middle-class US citizens'' \cite{costanza2020design}. It can make more visible the unequal distribution of benefits and harms in society as well as highlight the role algorithms can play toward either reinforcing or dismantling those relationships \cite{abebe2020roles}.

\subsection*{Shortest Paths for Seam Carving}

Seam carving is an approach for content-aware image resizing \cite{avidan2007seam}. Hug (in \citeauthor{parlante2015nifty}) describe \emph{Seam Carving} as an assignment for Basic Data Structures. Rather than shrinking images by scaling the image or cropping-out the edges, seam carving removes the least-noticeable vertical or horizontal \emph{seam}, or path of pixels from top-to-bottom or left-to-right.

Shortest paths algorithms can find a least-noticeable seam by representing the image as a graph where vertices represent pixels and edge weights represent the visual difference between adjacent pixels as defined by an \emph{energy function}. Since the energy function defines the visual difference between pixels, it determines the seams that are selected by the shortest paths algorithm and then ultimately removed. In their SIGGRAPH presentation video, \citeauthor{avidan2007seam} compared different energy functions on an image of a dark-skinned woman, including 2 functions that erased body parts. Although they recognize how the ``results depend on the given image,'' an evaluation of the role of skin color and gender \cite{buolamwini2018gender} is notably absent.

Affordance analysis not only attends to the information erased by abstractions as we've seen in the preceding case studies, but also extends critique to the \emph{cultural and epistemic abstractions} of CS education that lead to the production of anti-political subjects that ``acknowledge ethical and political dimensions'' but ``divest them from `what counts' as [CS]'' \cite{malazita2019infrastructures}.

\section*{Discussion}

Affordance analysis problematizes the study of data structures and algorithms by centering ethical dilemmas. Computational abstractions make certain technical solutions more accessible and likely than other solutions. Since these abstractions embody affordances, and affordances embody values, the choice of abstraction can lead to differential consequences. Affordance analysis advances beyond ``deep tech ethics'' \cite{ferreira2021deep} by implicating the design of abstractions in the outcomes that they produce. Peck (in \citeauthor{doore2020assignments}) suggests that this experience of design, evaluation, and critique can lead to ethical reflections: ``What does it mean to design a fair algorithm? What is the human cost of efficiency? What systemic advantages/disadvantages are our algorithms [likely] to amplify?''

Affordance analysis offers a more critical algorithm analysis, one that centers the rightful presence of political struggles by examining how sociotechnical systems distribute benefits and harms. Each of the 5 case studies highlighted the implicit and explicit ways that algorithms structure and reinforce social relations. By teaching a more critical algorithm analysis, we not only teach the limits of computation and data \cite{ko2020time}, but also support a greater appreciation, understanding, and responsibility toward equity---a key element of cultural competence \cite{washington2020twice}.

Although affordance analysis considers social relations, it is inherently limited to the algorithmic components of a sociotechnical system. Affordance analysis recognizes and critiques how abstractions embody values, but it provides less direction for redressing design values and limitations. \citeauthor{costanza2020design} defines design justice as a frame for rethinking the ``universalizing assumptions behind affordance theory'' and asking ``questions about how inequality structures affordance perceptibility and availability.'' To move beyond universalizing assumptions, affordance analysis must be considered in dialogue with ecological concerns over how data structures and algorithms are developed: the design practices, design narratives, design sites, and design pedagogies that create contemporary social conditions \cite{costanza2020design}. For example, to rethink design practices, rather than identify a `better' prioritization for content moderation, we might instead collaborate with users (intended beneficiaries), content moderators (local experts), and governments (regulatory institutions) to design a more sustainable and restorative approach to moderating user-generated content. We might expect social media platforms to treat content moderation as a primary rather than tertiary concern \cite{abebe2020roles}.

Affordance analysis is not mutually exclusive with traditional learning objectives for Basic Data Structures such as asymptotic analysis \cite{porter2018developing}. In fact, we consider it important to study both of these framings ``in tandem'' \cite{kafai2019theory} in order to recognize the various past, present, and future purposes and limits of CS education \cite{vakil2018ethics}. Asymptotic analysis can also be a critical algorithm analysis: the demand for efficient algorithms and computation centralizes hegemonic power in technosocial elites. But it is important to recognize that, historically, CS education has sidelined justice-centered framings in favor of cognitive framings focused on developing students' knowledge, skills, and understanding of CS concepts and practices at the cost of developing students' disciplinary identities \cite{vakil2018ethics, vakil2020ve, agarwal2019integrating}.

Affordance analysis expands the definition of algorithm analysis by foregrounding the affordances embodied within computational abstractions. As a step toward a more critical CS education, affordance analysis makes space for the rightful presence of political struggles in the computing classroom \cite{calabrese2020beyond}.

\printbibliography

\end{document}